\documentclass[12pt]{article}

\usepackage{graphics,amssymb,epsfig,float}
\usepackage[usenames,dvips]{color}
\usepackage{graphicx}
\usepackage{epsfig}
\usepackage{rotating}
\usepackage{dcolumn}
\usepackage{bm}
\usepackage{cite}
\usepackage{amsmath}
\usepackage{setspace}

\def\lsim{\;\raise0.3ex\hbox{$<$\kern-0.75em\raise-1.1ex\hbox{$\sim$}}\;}
\def\gsim{\;\raise0.3ex\hbox{$>$\kern-0.75em\raise-1.1ex\hbox{$\sim$}}\;}
\def\beq{\begin{equation}}   \def\eeq{\end{equation}}
\def\ba{\begin{array}}       \def\ea{\end{array}}
\def\bea{\begin{eqnarray}}   \def\eea{\end{eqnarray}}

\begin{document}

\begin{titlepage}
\begin{flushright}
LPT Orsay 12-21\\
PCCF RI 12-01
\end{flushright}

\begin{center}
\vspace{1cm}
{\Large\bf Modified Signals for Supersymmetry in the NMSSM
with a Singlino-like LSP}
\vspace{2cm}

{\bf{Debottam Das$^1$, Ulrich Ellwanger$^1$ and Ana M. Teixeira$^2$}}
\vspace{1cm}\\
\it $^1$  Laboratoire de Physique Th\'eorique, UMR 8627, CNRS and
Universit\'e de Paris--Sud,\\
\it B\^at. 210, 91405 Orsay, France \\
\it $^2$ Laboratoire de Physique Corpusculaire, CNRS/IN2P3 -- UMR
6533,\\
\it Campus des C\'ezeaux, 24 Av. des Landais, F-63171 Aubi\`ere Cedex,
France\\

\end{center}

\vspace{1cm}

\begin{abstract}
In the framework of the NMSSM with a singlino-like LSP, 
we study quantitatively the impact of the additional bino~$\to$~singlino
cascade on the efficiencies in
several search channels for supersymmetry of the ATLAS and CMS
collaborations. Compared to the MSSM, the additional cascade reduces the
missing transverse energy, but leads to additional jets or leptons. For
the NMSSM benchmark lines which generalize cMSSM benchmark points, the
efficiencies in the most
relevant 2/3~jet + missing energy search channels can drop by factors
$\sim 1/3$ to $\sim 1/7$, and can reduce the present lower bounds on
$M_{1/2}$ by as much as $\sim 0.9\ - 0.75$ in the NMSSM for large
bino--singlino mass differences. The larger efficiencies in multijet or
multilepton search channels are not strong enough to affect this
conclusion. In the fully constrained cNMSSM, sparticle decay cascades
via the lightest stau can lead to signal cross sections in multilepton
and $2\,\tau$ search channels which are potentially visible at the LHC
with 7~TeV center of mass energy.
\end{abstract}

\end{titlepage}

\newpage
\section{Introduction}

The Next-to-Minimal Supersymmetric Standard Model (NMSSM
\cite{Ellwanger:2009dp}) is the simplest super\-symmetric (SUSY)
extension of the Standard Model with a scale invariant superpotential,
i.e. where the soft SUSY breaking terms are the only dimensionful
parameters. A supersymmetric Higgs mass term $\mu$, as required in the
MSSM, is generated dynamically by a vacuum expectation value (vev) of a
gauge singlet (super-)field $S$, and is automatically of the order of
the supersymmetry breaking scale. The attractive features of the MSSM
are preserved, like a solution of the hierarchy problem, the unification
of the running gauge coupling constants at a Grand Unification (GUT)
scale, and a dark matter candidate in the form of a stable lightest SUSY
particle (LSP).

Using data from $1-2$~fb$^{-1}$ of integrated luminosity at the LHC at
7~TeV center of mass (c.m.) energy, searches for supersymmetry by the
ATLAS  \cite{Aad:2011ib,ATLAS:2011ad,Aad:2011qa,1110:6189,
1111:4116,:2011cw} and CMS \cite{Chatrchyan:2011zy,CMS11-004,
CMS11-005,CMS11-007,CMS11-008,CMS11-009, CMS11-010,CMS11-011,CMS11-013,
CMS11-015,Khachatryan:2011tk} collaborations have not led to the discovery of signals of
supersymmetric particles. 

However, already within the MSSM, the interpretation of the absence (or
presence) of signals in the many possible channels depends on the soft
SUSY breaking terms: Assuming R-parity conservation and a
neutralino-like LSP, missing transverse energy $E_T^\text{miss}$ together
with high $p_T$ jets and/or leptons are used as search criteria. These
signatures depend on the decay cascades of the initially produced
squarks and gluinos (dominant at a hadron collider), which depend on the
spectrum and the couplings of sparticles (squarks, gluino, sleptons,
charginos and neutralinos), and hence on the soft SUSY breaking terms.

A popular choice for the soft SUSY breaking terms is the constrained
MSSM (cMSSM) where scalar masses $m_0$, gaugino masses $M_{1/2}$ and
trilinear couplings $A_0$ are assumed to be universal at the GUT scale. 
In the cMSSM, the lightest neutralino $\chi^0_1$ is typically the LSP, and
bino-like in most of the parameter space (apart from very small values
of $m_0$, or very large values of both $m_0$ and $M_{1/2}$). $\chi^0_1$
appears as one of the final states in every sparticle decay cascade, and
is responsible for the missing transverse energy.

Assuming fixed values for $A_0$ and $\tan\beta$ (the ratio of the two
Higgs vevs $\left<H_u\right>/\left<H_d\right>$), the ATLAS and CMS
collaborations have deduced lower bounds on $m_0$ and $M_{1/2}$ in the
$m_0 - M_{1/2}$~plane in the cMSSM. Recently, the absence of signals has
been interpreted within more general scenarios of soft SUSY breaking
terms as the phenomenological MSSM \cite{Sekmen:2011cz, Arbey:2011un,
Papucci:2011wy} and anomaly mediation \cite{Allanach:2011qr}. In the
present paper, we perform a first study of possible modifications of
signals for supersymmetric particles in the framework of a
generalization of the cMSSM towards the NMSSM.

The additional gauge singlet superfield $\hat S$ in the NMSSM leads to
additional physical states in the CP-even and CP-odd Higgs sectors, as
well as an additional singlino-like neutralino. The impact of the
additional Higgs states can be very important for Higgs searches (see
\cite{Ellwanger:2011sk} for a review), but in the present paper we
concentrate on the possible impact of the additional singlino-like
neutralino on searches for supersymmetry. This becomes par\-ti\-cu\-larly
relevant in regions of the NMSSM parameter space where the singlino-like
neutralino is the LSP $\chi^0_1$: Then the singlino-like neutralino
appears as one of the final states in every sparticle decay cascade.

However, the couplings of the singlino-like neutralino to all other
sparticles are typically very small. Then the branching ratios for
decays of all sparticles into the singlino-like LSP $\chi^0_1$ are
small, with the exception of the NLSP which can decay only into
$\chi^0_1$ if R-parity is conserved. Hence sparticle decay cascades will
evolve as in the MSSM, with an additional final decay
\beq\label{eq:1}
\chi^0_2 \to \chi^0_1 + X
\eeq
(if the next-to-lightest neutralino $\chi^0_2$ is the next-to-LSP
(NLSP) which is, however, not necessarily the case in the NMSSM).
Depending on the mass difference
\beq\label{eq:2}
\Delta_M = M_{\chi^0_2} - M_{\chi^0_1}\,,
\eeq
and on the nature of the decay products $X$ in (\ref{eq:1}), this additional decay of
the NLSP will generally reduce $E_T^\text{miss}$, but will lead to more jets
and/or leptons. Hence it affects all signatures used in searches for
supersymmetry. A priori, it is not clear how the reduction of
$E_T^\text{miss}$ and the additional jets and/or leptons affect the
efficiencies in the various SUSY search channels. A quantitative
analysis of the modifications of the efficiencies in the NMSSM, for some
relevant search channels for supersymmetry (without and with leptons),
is the purpose of the present paper.

Like in the general MSSM, many different scenarios for the soft SUSY
breaking terms are possible in the NMSSM. We found it useful to compare
NMSSM and MSSM scenarios which are as close as possible: Similar squark,
slepton and gaugino spectra, but with an additional singlino-like LSP in
the NMSSM. For convenience, we compare a semi-constrained version of the
NMSSM (sNMSSM) to the cMSSM: We choose the same values for $m_0$,
$M_{1/2}$, $A_0$ and $\tan\beta$ as for some benchmark points of the
cMSSM, but allow for non-universal singlet-specific soft SUSY breaking
terms $m_S^2$, $A_\lambda$ and $A_\kappa$ (see the next section) such
that the singlino-like neutralino is the LSP $\chi^0_1$, and the
extended Higgs sector is in agreement with constraints from LEP.

The advantage of this approach is that, for each choice of $m_0$,
$M_{1/2}$, $A_0$ and $\tan\beta$, we can compare directly the
efficiencies in various channels (after appropriate cuts) between the
cMSSM and the sNMSSM. This allows to estimate to which extent the
additional singlino-like neutralino affects present or future limits or
excesses in the $m_0 - M_{1/2}$~plane in comparison to the cMSSM as a
function of $\Delta_M$, without the need for a novel analysis of
backgrounds and systematic errors.

Since we cannot study the complete $m_0 - M_{1/2}$~plane, we consider
three benchmark points of the cMSSM as defined in
\cite{AbdusSalam:2011fc}, with not too large values of $m_0$ and
$M_{1/2}$ (presently not excluded, but of possible future relevance for
the LHC at 7~TeV c.m.~energy). For each point we vary the NMSSM-specific
parameters in a domain where the singlino-like neutralino is the LSP
($\chi^0_1$), allowing $\Delta_M$ to vary within the largest possible
range complying with constraints from LEP2 on the Higgs sector. Thus
each benchmark point of the cMSSM is promoted to a benchmark
line in the sNMSSM. We then study the ratios $R$ of efficiencies
sNMSSM/cMSSM for 5 search channels of the  ATLAS jets + missing
transverse momentum (0 leptons) analysis in \cite{Aad:2011ib}, and 4
search channels of the ATLAS multijet + missing transverse momentum (0
leptons) analysis in \cite{Aad:2011qa}, always as function of
$\Delta_M$. (For our purpose it is useful that the lower bounds in the
$m_0 - M_{1/2}$~plane are given for each search channel separately on
the web page \cite{ATLASweb}.)

Actually these ratios $R$ always tend towards~1 in the limit $\Delta_M
\to 0$: Then the complete (missing) energy is transferred from
$\chi^0_2$ to $\chi^0_1$ in the decay (\ref{eq:1}), and no energy
remains to generate high $p_T$ jets or leptons from the decay products
$X$. However, for larger $\Delta_M$ we find that, after cuts
corresponding to the most sensitive 2/3-jet, 0 lepton search channels,
the efficiencies in the sNMSSM can be smaller by a factor $\sim 1/7$ as
compared to the cMSSM! Simultaneously, in the less sensitive multijet
search channels the efficiencies in the sNMSSM would be larger than in
the cMSSM.

Since the additional decay products $X$ in (\ref{eq:1}) can consist of
leptons, it is also interesting to study search channels of the CMS
multilepton analysis in \cite{CMS11-013} for the sNMSSM. Here, however,
the corresponding cMSSM efficiencies are so small that it is not
meaningful to give ratios of efficiencies sNMSSM/cMSSM; hence we give
estimates for absolute signal cross sections (production cross sections
$\times$ efficiencies) for those values of $\Delta_M$ which correspond
to the largest efficiencies. The absolute signal cross sections allow to
estimate the future discovery potential in these channels.

The phenomenologically allowed region in the $m_0 - M_{1/2}$~plane in
the sNMSSM is actually somewhat larger than in the cMSSM: Small values
for $m_0$ would lead to a stau ($\tilde{\tau}_1$) LSP in the cMSSM,
while in the sNMSSM the singlino-like neutralino can be lighter than
the $\tilde{\tau}_1$ even for $m_0\to 0$. In particular, this scenario
is always realised in the fully constrained cNMSSM
\cite{arXiv:0803.0253,arXiv:0811.2699}, where the soft singlet mass term
$m_S$ satisfies $m_S = m_0$ at the GUT scale, and $m_S$ must be small in
order to allow for a non-vanishing singlet~vev. The sparticle decay
cascades in the cNMSSM are quite peculiar: Since the singlino-like LSP
couples only very weakly to all MSSM-like sparticles, the latter decay
first into the $\tilde{\tau}_1$ NLSP. Only subsequently does the
$\tilde{\tau}_1$ NLSP decay into the singlino-like LSP $\chi^0_1$,
leading finally to 2 $\tau$'s in the final state of each sparticle decay
chain. However, the $\tau$'s from the final $\tilde{\tau}_1 \to
\chi^0_1$ decay are always quite soft due to the small $\tilde{\tau}_1 -
\chi^0_1$ mass difference. Hence, only the more energetic $\tau$'s from
the sparticle~$\to \tilde{\tau}_1 + \tau$ decay constitute a visible
particular feature of the cNMSSM. (In the cMSSM in the
stau-coannihilation region, $\tau$ production from decays into and of
the $\tilde{\tau}_1$ NLSP can also be expected, although not as frequently
as in the cNMSSM.)

For the LHC at 14~TeV c.m.~energy, appropriate cuts for searches for the
cNMSSM (and ways to distinguish it from the cMSSM) have been proposed
and studied in \cite{Ellwanger:2010es}. Using data from $1$~fb$^{-1}$ of
integrated luminosity at the LHC at 7~TeV c.m. energy, $2\,\tau$ channels
have been analysed by CMS in \cite{CMS11-007}.

In what follows, we also perform some analyses of the two cNMSSM 
benchmark points with the lowest values of $M_{1/2}$, defined in
\cite{AbdusSalam:2011fc}. (LEP bounds on the Higgs sector imply $M_{1/2}
\gsim 520$~GeV in the cNMSSM
\cite{arXiv:0803.0253,arXiv:0811.2699}.) As done for the comparison of the
sNMSSM with the cMSSM, we study the ratios of efficiencies for 5 search
channels of the  ATLAS jets + missing transverse momentum (0 leptons)
analysis in \cite{Aad:2011ib} (using similar realistic cMSSM points with
small, but non-vanishing values of $m_0$, which are not excluded by
present searches). In these search channels, the efficiencies in the
cNMSSM and the cMSSM turn out to be quite similar. In addition we give
estimates for absolute signal cross sections for 4 search channels of
the CMS multilepton analysis in \cite{CMS11-013} and the CMS $2\,\tau$
channels \cite{CMS11-007}. The corresponding efficiencies in the cMSSM
would be very small, which could allow to distinguish these models in
the future.

The remainder of the paper is organised as follows: In the next section
we discuss the NMSSM with a singlino-like neutralino, define 3 different
benchmark lines of the sNMSSM, and 2 benchmark points of the cNMSSM. In
Section~3 we describe the tools used for our Monte Carlo study, and the
cuts used for the ATLAS/CMS search channels. Section~4 contains our main
results. For each of the 3 benchmark lines of the sNMSSM, we present
first the branching ratios for the decay (\ref{eq:1}) into the
additional final states $X$ as function of $\Delta_M$. We then give the
ratios of efficiencies sNMSSM/cMSSM for different supersymmetry search
channels used by ATLAS, and estimates for the signal cross sections for
4 search channels of the CMS multilepton analysis for values of
$\Delta_M$ corresponding to the largest efficiencies. For the cNMSSM
points we provide, in addition, estimates for absolute signal cross
sections for the CMS $2\,\tau$ channels. Conclusions and an outlook are
presented in Section~5.

\section{The NMSSM with a Singlino-like LSP}

The NMSSM differs from the MSSM due to the presence of the gauge singlet
superfield $\hat S$. In the simplest realisation of the NMSSM, the Higgs mass
term $\mu \hat H_u \hat H_d$ in the MSSM superpotential $W_{MSSM}$ 
is replaced by the coupling $\lambda$ of $\hat S$ to $\hat H_u$ and
$\hat H_d$, and a self-coupling $\kappa \hat S^3$.  Hence, in this version
the superpotential $W_{NMSSM}$ is scale invariant, and given by:
\beq\label{eq:3}
W_{NMSSM} = \lambda \hat S \hat H_u\cdot \hat H_d + \frac{\kappa}{3} 
\hat S^3 +\dots\; ,
\eeq
where the dots denote the Yukawa couplings of $\hat H_u$ and $\hat H_d$
to the quarks and leptons as in the MSSM. Once the scalar component of
$\hat S$ develops a vev $s$, the first term in $W_{NMSSM}$ generates an
effective $\mu$-term with
\beq\label{eq:4}
\mu_\mathrm{eff}=\lambda s\; .
\eeq

The NMSSM-specific soft SUSY breaking terms consist of a mass term for
the scalar components of $\hat S$, and trilinear interactions associated
to the terms in $W_{NMSSM}$:
\beq\label{eq:5}
 -{\cal L}_{NMSSM}^{Soft} = m_{S}^2 |S|^2 +\Bigl( \lambda A_\lambda\,
H_u \cdot H_d \,S +  \frac{1}{3} \kappa  A_\kappa\,  S^3 \Bigl)\ +\
\mathrm{h.c.}\;.
\eeq

The neutral CP-even Higgs sector contains 3 states $H_i$, which are
mixtures of the CP-even components of the superfields $\hat H_u$, $\hat
H_d$ and $\hat S$. Their masses are described by a $3 \times 3$ mass
matrix ${\cal M}^2_{H\,ij}$. The neutral CP-odd Higgs sector contains 2
physical states $A_i$, whose masses are described by a $2 \times 2$ mass
matrix ${\cal M}^2_{A\,ij}$. In the neutralino sector we have 5 states
${\chi}^0_i$, which are mixtures of the bino $\tilde{B}$, the neutral
wino $\tilde{W}^3$, the neutral higgsinos from the superfields $\hat
H_u$ and $\hat H_d$, and the singlino from the superfield $\hat S$.
Their masses are described by a $5 \times 5$ mass matrix ${\cal
M}_{\chi^0\,ij}$.

Subsequently, it is of interest to consider the singlet-like
components of these mass matrices (given in \cite{Ellwanger:2009dp}),
for simplicity in the typical range $s \gg v_u,\, v_d$, where $v_u,\,
v_d$ are the vevs of $H_u$, $H_d$:
\beq\label{eq:6}
{\cal M}^2_{H\,33} \sim \kappa s\left(A_\kappa + 4 \kappa s\right)\; ,
\eeq
\beq\label{eq:7}
{\cal M}^2_{A\,22} \sim -3\kappa s A_\kappa\; ,
\eeq
\beq\label{eq:8}
{\cal M}_{\chi^0\,55} = 2 \kappa s\; .
\eeq
From the above one easily derives
\beq\label{eq:9}
{\cal M}^2_{\chi^0\,55} \sim {\cal M}^2_{H\,33} + \frac{1}{3}
{\cal M}^2_{A\,22}\; .
\eeq
Since both matrix elements ${\cal M}^2_{H\,33}$ and ${\cal M}^2_{A\,22}$
must be positive, one can conclude that none of them can be large
if ${\cal M}_{\chi^0\,55}$ is small, and notably that ${\cal
M}^2_{H\,33} < {\cal M}^2_{\chi^0\,55}$. 

In general, these matrix elements differ from the physical masses due to
mixing effects. However, the mixing angles are small for small
off-diagonal singlet-doublet matrix elements (which are proportional to
$\lambda$) or large mass differences, and for not too large $\lambda$
the physical masses in the singlet sector are quite close to the above
expressions. Hence, a light singlet-like neutralino is always
accompanied by a lighter singlet-like CP-even Higgs boson, and the
singlet-like CP-odd Higgs boson is maximally $\sim \sqrt{3}$ times as
heavy as the singlet-like neutralino, but typically lighter. This has
important consequences for the possible final states $X$ in the decay
(\ref{eq:1}), where these Higgs states are often kinematically allowed
and constitute possible 2-body decay channels ($X \equiv H_S$ or $X
\equiv A_S$ where the index $S$ denotes a mostly singlet-like state) of
the bino-like neutralino~${\chi}^0_2$.

These considerations are valid for the general NMSSM with a scale
invariant superpotential. As stated in the Introduction, we consider
subsequently the semi-constrained sNMSSM where the non-singlet scalar
masses, non-singlet trilinear couplings and gaugino masses are universal
at the GUT scale with values denoted by
$m_0$, $A_0$ and $M_{1/2}$, respectively. The remaining parameters
$\lambda$, $\kappa$, $m_S^2$, $A_\lambda$ and $A_\kappa$ of the sNMSSM
are chosen as follows: First, we choose a small value for $\lambda$,
implying that the singlet-like Higgs bosons and
the singlino-like neutralino couple only weakly to all other particles
and sparticles. Then the light singlet-like CP-even Higgs boson is
compatible with LEP constraints (due to its small coupling to the $Z$
boson), and the choice of the sNMSSM specific parameters has little
impact on the MSSM-like Higgs and sparticle spectrum. This ensures that
the differences of the efficiencies with respect to the cMSSM are only
due to the presence of the singlino-like neutralino, and not due to
modifications of e.g. the higgsino and/or Higgs spectra: The MSSM-like
parameters $\mu$ and $M_A$ are kept fixed, which determines implicitely
the values of $m_S^2$ and $A_\lambda$ in the sNMSSM. (Fixing
$\mu_\mathrm{eff}$ for fixed $\lambda$ implies from (\ref{eq:4}) that
the vev $s$ is fixed as well.) Then we vary $\kappa$ and $A_\kappa$ such
that the singlino-like neutralino mass (\ref{eq:8}) is below the
bino-like neutralino mass, and the matrix elements (\ref{eq:6}) and
(\ref{eq:7}) remain positive. Positive singlet-like Higgs masses
compatible with LEP constraints imply actually lower bounds on the
singlino-like neutralino mass, but these
lower bounds still allow for $\Delta_M$ to vary over a wide range.

As in the cMSSM, the dark matter relic density does not generally comply
with the WMAP bounds for generic points in the sNMSSM $m_0 -
M_{1/2}$~plane, being too large for a singlino-like LSP.
As in \cite{AbdusSalam:2011fc}, one could assume a deviation from
standard Big-Bang cosmology to reduce the relic density, or a small
R-parity violation that renders the LSP unstable. Alternatively, one
could modify the parameters in the Higgs sector (and choose a larger
value of $\lambda$), such that a CP-even or CP-odd s-channel resonance
is available for LSP pair annihilation. Such modifications would have
little impact on our subsequent results.

As stated in the Introduction, we define three benchmark lines in the
sNMSSM for which the values for $m_0$, $M_{1/2}$, $A_0$ and $\tan\beta$
correspond to three benchmark points in the cMSSM defined in
\cite{AbdusSalam:2011fc}: 10.1.1, 10.4.1 and 40.2.1. These
correspond to different (but not too large) values of $m_0$ and
$M_{1/2}$, with the first number (10 or 40) denoting the value of
$\tan\beta$. In each case we fix a small value of $\lambda$, use
the cMSSM values for $\mu_\mathrm{eff}$ and $M_A$, and vary $\kappa$ and
$A_\kappa$ such that $\Delta_M = M_{\chi^0_2} - M_{\chi^0_1}$ varies
from 0 to a maximal value determined by LEP constraints on the Higgs
sector. The singlet-like CP-even and CP-odd Higgs masses vary somewhat
with varying $\kappa$ and $A_\kappa$, but are always small as shown
above. All MSSM-like sparticle properties (notably the squark and gluino
masses) are practically constant along the sNMSSM benchmark lines.
Hence the sparticle production cross sections in the sNMSSM remain the
same as in the cMSSM by construction; these are not affected by the
additional singlino-like neutralino (or Higgs) states.

In Table~\ref{tab:1} we indicate the most relevant properties of the
three benchmark lines: First the cMSSM-like parameters $m_0$, $M_{1/2}$,
$A_0$, $\tan\beta$, $\mu_\mathrm{eff}$ and $M_A$, and the gluino mass
$m_{\tilde{g}}$, the average squark masses $\left< m_{sq} \right>$ of
the first families, and the bino-like neutralino mass $M_{\chi^0_2}$
(note that $\chi^0_2$ is the NLSP in the sNMSSM). Subsequently we give
the ranges of the NMSSM-specific parameters, the singlino-like
neutralino mass  $M_{\chi^0_1}$ and the SM-like and singlet-like Higgs
masses $M_{H_{SM}}$, $M_{H_S}$ and $M_{A_S}$, respectively.
($M_{H_{SM}}$ can be larger or smaller than $M_{H_S}$. The
sparticle and Higgs spectrum is obtained with the help of the code
NMSPEC \cite{Ellwanger:2006rn} within the version 3.0.2 of NMSSMTools
\cite{Ellwanger:2004xm, Ellwanger:2005dv}.) Finally we provide the total
sparticle production cross section $\sigma_\mathrm{Tot}$ at the
7~TeV~LHC as obtained by Prospino (at next-to-leading order)
\cite{Beenakker:1996ch,Beenakker:1996ed, Beenakker:1999xh}.

\begin{table}[h!]
\begin{center}
\begin{tabular}{|c|c|c|c|} \hline
Point:                  & 10.1.1 & 10.4.1  & 40.2.1  \\\hline
$M_{1/2}$               & 500  & 350 & 450  \\\hline
$m_0$                   & 125  & 750 & 550  \\\hline
$A_0$                   & 0    & 0   &$-500$  \\\hline
$\tan\beta$             & 10   & 10  & 40  \\\hline
$\mu_\mathrm{eff}$      & 635 & 465  & 645 \\\hline
$M_A$                   & 720 & 895  & 710 \\\hline
$m_{\tilde{g}}$         & 1145 & 870 & 1065 \\\hline
$\left< m_{sq} \right>$ & 1030 & 1040&1080\\\hline
$M_{\chi^0_2}$          & 205  & 143  & 187  \\\hline \hline
$\lambda$               &$10^{-3}$& 0.013  &  $10^{-3}$ \\\hline
$\kappa$                &$-1.6\cdot 10^{-4}\ {\dots}\ -2\cdot 10^{-5}$
                        &$-2\cdot 10^{-3}\ {\dots}\ -8.7\cdot 10^{-4}$
			&$1.8\cdot 10^{-5}\ \dots\ 1.4\cdot 10^{-4}$
			\\\hline
$A_\kappa$              &0.7\ \dots\ 1.6
                        &$0 \ {\dots}\ 150$   
                        &$-7\ {\dots}\ -4.2$\\\hline
$M_{\chi^0_1}$          &$25\ {\dots}\ 205$
                        &$50\ {\dots}\ 143$   
			&$23\ {\dots}\ 187$    \\\hline
$M_{H_{SM}}$            &$\sim 115$
                        &   $ 115\ {\dots}\ 117$
                        &$\sim  117$   
\\\hline
$M_{H_S}$               &$ 25\ {\dots}\ 205$   & $ 55\ {\dots}\ 89$   
&$ 21\ {\dots}\ 186$   \\\hline
$M_{A_S}$               &$8\ {\dots}\ 20$   &  $ 5\ {\dots}\ 160$
 &$6\ {\dots}\ 34$
\\\hline
$\sigma_\mathrm{Tot}$& 82~fb & 300~fb & 87~fb \\\hline
\end{tabular}
\end{center}
\caption{Parameters, some sparticle and Higgs masses, and the total
sparticle production cross section for three sNMSSM benchmark lines
corresponding to the cMSSM benchmark points 10.1.1, 10.4.1 and 40.2.1
from \cite{AbdusSalam:2011fc} (masses in GeV, rounded to 5~GeV accuracy
except for $M_{\chi^0_2}$ and the Higgs masses).}
\label{tab:1}
\end{table}

In addition, we study two points of the fully constrained cNMSSM
\cite{arXiv:0803.0253,arXiv:0811.2699} where the singlet-specific soft
SUSY breaking terms $m_S$, $A_\lambda$ and $A_\kappa$ are also
respectively given by $m_0$, $A_0$ at the GUT scale, and the
$\tilde{\tau}_1$ is the NLSP. To comply with a dark matter relic density
compatible with WMAP constraints, the singlino-like $\chi^0_1$ must be a
few GeV lighter than the $\tilde{\tau}_1$ which defines a nearly unique
line in the $M_{1/2}$, $A_0$, $\tan\beta$ parameter space
\cite{arXiv:0803.0253,arXiv:0811.2699} (taking $m_0=0$, $M_{1/2} >
520$~GeV and $\lambda=10^{-3}$, such that the CP-even Higgs sector
complies with LEP constraints).

In \cite{AbdusSalam:2011fc}, benchmark points cNMSSM.1 and cNMSSM.2
(amongst others) have been defined. In the cMSSM parameter space,
similar points can be found in the so-called stau-coannihilation region:
For identical values of $M_{1/2}$, $A_0$ and $\tan\beta$, it suffices to
choose small non-vanishing values for $m_0$ such that the
$\tilde{\tau}_1$ mass is just above the bino mass, which leads again to
a good relic density and a sparticle spectrum which is otherwise very
close to the cNMSSM. We found it appropriate to compare efficiencies in
the mostly used jets + missing energy channels for the points cNMSSM.1
and cNMSSM.2 to efficiencies for similar points in the cMSSM, denoted
here by cMSSM.1 and cMSSM.2.
In Table~\ref{tab:2} we give the most relevant properties of these
benchmark points cNMSSM.1 and cNMSSM.2, together with the points
cMSSM.1 and cMSSM.2 for which $m_0$ and hence the squark/slepton masses
are slightly larger. (For the cNMSSM points $\chi^0_1$ is singlino-like;
for the corresponding cMSSM points, $\chi^0_1$ is mostly bino-like.)

\begin{table}[h!]
\begin{center}
\begin{tabular}{|c|c|c|c|} \hline
Point:                  & cNMSSM.1 & cNMSSM.2   \\\hline
$M_{1/2}$               & 520      & 600   \\\hline
$m_0$                   & 0        & 0   \\\hline
$A_0$                   & -146.5   & -171     \\\hline
$\tan\beta$             & 22.2     & 23.3    \\\hline
$\lambda$               &$10^{-3}$ & $10^{-3}$ \\\hline
$\kappa$                &$1.1\cdot 10^{-4}$&$1.1\cdot 10^{-4}$\\\hline
$m_{\tilde{g}}$         & 1190     & 1360  \\\hline
$\left< m_{sq} \right>$ & 1060     & 1200\\\hline
$M_{\chi^0_1}$          & 146.4    & 171  \\\hline
$M_{\tilde{\tau}_1}$    & 150.5    & 174.5  \\\hline
$M_{H_{SM}}$            & 114      & 115     \\\hline
$M_{H_S}$          & 103      & 121    \\\hline
$M_{A_S}$               & 179      & 209    \\\hline \hline
 &cMSSM.1 &cMSSM.2\\\hline
$m_0$                   & 170        & 194   \\\hline
$M_{\chi^0_1}$          & 214.8    & 249.8  \\\hline
$M_{\tilde{\tau}_1}$    & 221.6      &254.1  \\\hline
$\sigma_\mathrm{Tot}$   & 73~fb    & 28~fb  \\\hline
\end{tabular}
\end{center}
\caption{Parameters, some sparticle and Higgs masses, and the total
sparticle production cross section for two cNMSSM benchmark points and
nearby cMSSM points in the stau co-annihilation region, all of which
account for a good dark matter relic density.}
\label{tab:2}
\end{table}

\section{Monte Carlo Simulations and Search Channels}

For the calculation of the matrix elements we use MadGraph/MadEvent~5
\cite{Alwall:2011uj}, which includes Pythia~6.4 \cite{Sjostrand:2006za}
for showering and hadronisation. Matching of the differential jet cross
sections is performed according to the prescriptions in
\cite{Alwall:2008qv}. The sparticle branching ratios are obtained with
the help of the code NMSDECAY \cite{Das:2011dg} (based on SDECAY
\cite{Muhlleitner:2003vg}), and are passed to Pythia.

The output is given in StdHEP-format to the fast detector simulation
Delphes \cite{Ovyn:2009tx}. Inside Delphes, the appropriate ATLAS or CMS
detector cards are used, together with the appropriate jet
reconstruction algorithm. The jet reconstruction is performed with
FastJet \cite{Cacciari:2005hq}.

Subsequently we apply cuts corresponding to the following search
channels S1~--~S4:
\begin{itemize}

\item{S1:} ATLAS jets + missing transverse momentum (0 leptons) analysis
\cite{Aad:2011ib}. 5 different signal regions are defined in Table~1 in
\cite{Aad:2011ib}. The cuts on $E_T^\mathrm{miss}$ and $p_T$ of the
leading jet are always $> 130$~GeV. 4 signal regions will be denoted by
$2j$, $3j$, $4j$ and $4jL$, where the $p_T$ of the second, third and
fourth jet satisfy always $p_T > 40$~GeV, and $m_\mathrm{eff} >
1000$~GeV for $2j$, $3j$, $4j$, but $m_\mathrm{eff} > 500$~GeV for
$4jL$. In a fifth ``high mass'' region, denoted by $4jH$, the second,
third and fourth jet satisfy always $p_T > 80$~GeV, and $m_\mathrm{eff}
> 1100$~GeV (see \cite{Aad:2011ib} for more details).

\item{S2:} ATLAS multijet + missing transverse momentum (0 leptons)
analysis \cite{Aad:2011qa}. Here 4 signal regions are denoted by $7j55$,
$8j55$, $6j80$ and $7j80$, where the numbers after the jet
multiplicities denote the lower cut on the jet $p_T$ (see
\cite{Aad:2011qa} for more details on the event selection).

\item{S3:} CMS analysis of multilepton signatures \cite{CMS11-013}.
Numerous different search channels have been considered in
\cite{CMS11-013}, depending on the lepton number, lepton species and
charges. In addition, either a cut on $E_T^\mathrm{miss}$ (MET) $>
50$~GeV, or a cut on $H_T > 200$~GeV was applied ($H_T$ is defined as
the scalar sum of the transverse jet energies for all jets with $E_T >
40$~GeV). In some channels, excesses of events w.r.t. the SM have been
observed in \cite{CMS11-013}, but the event rates are still too small
(and the systematical/statistical errors too large) to consider
these excesses as significant.

Subsequently we sum over all leptons including (hadronically decaying)
$\tau$ leptons, and distinguish only the search channels MET3, MET4, HT3
and HT4, where the numbers after MET or HT denote the number of leptons
including $\tau$'s. These search channels correspond to the lines $\sum
l (l/\tau)(l/\tau)$ and $\sum ll (l/\tau)(l/\tau)$ in Tables~1 and 2 in
\cite{CMS11-013}, where more details on the event selection can be
found.

\item{S4:} $2\,\tau$ search by the CMS collaboration \cite{CMS11-007},
where three signal regions have been defined. The first two require the
presence of a lepton and a hadronically decaying $\tau$ ($\tau_h$) of
opposite charge and $p_T > 20$~GeV, and two jets with $p_T > 30$~GeV. In
the first signal region denoted as $e/\mu\,\tau_h$~high~$E_T^\text{miss}$,
one requires $E_T^\text{miss} > 200$~GeV and $H_T > 300$~GeV. In the second
signal region denoted as $e/\mu\,\tau_h$~high~$H_T$, one requires $H_T >
400$~GeV and $E_T^\text{miss} > 150$~GeV. In a third signal region denoted as
$\tau_h\,\tau_h$, two hadronically decaying taus with $p_T > 15$~GeV,
two jets with $p_T > 100$~GeV and  \mbox{$\slash\hspace*{-3mm}{H}_T >
200$~GeV} are required. (\mbox{$\slash\hspace*{-3mm}{H}_T$} is defined
as \, \mbox{$\slash\hspace*{-3mm}{H}_T = |\sum_i \vec{p^i}_T|$}, where
the sum runs over all jets with $p_T > 30$~GeV.) This search channel is
of relevance only for the cNMSSM.

\end{itemize}

In all cases we compared our event rates to those of the MSSM benchmark
points given in these publications
\cite{Aad:2011ib,Aad:2011qa,CMS11-013,CMS11-007}, and found agreement
within $20\% - 30\%$. For each of the $\sim 80$ points studied in
Section~4, we simulated at least $10^4$ events.

Clearly, our estimates of the efficiencies in the
various channels are not as precise as the ones performed by the ATLAS
and CMS collaborations. Hence, in the cases S1 and S2 we analyse ratios
of efficiencies of benchmark lines 10.1.1, 10.4.1 and 40.2.1 of the
sNMSSM to the corresponding benchmark points in the cMSSM with the same
sparticle spectrum (except for the singlino-like LSP, see Section~2). We
can expect that systematic errors in our estimate of the efficiencies
in the various channels cancel to a large extent in such ratios.
Moreover, this procedure allows to translate bounds on $M_{1/2}$
obtained within the cMSSM into the parameter space of the NMSSM with a
singlino-like LSP, as we will discuss in Section~4.

This strategy fails for the multilepton analysis S3 (and the $2\,\tau$
analysis S4) where, in spite of the sum over different lepton species and
charges, the efficiencies in the cMSSM are so small that they are of the
order of our statistical errors. Therefore it is not appropriate
to define sNMSSM/cMSSM ratios, and
we will give estimates of the signal cross sections $\sigma$
after the event selection S3 for values of $\Delta_M$ which correspond
to the largest efficiencies. These allow to estimate the signal rates
for present and future luminosities.

\section{Results}

In this section we show our results for the ratios of efficiencies $R$ =
sNMSSM/cMSSM concerning the search channels S1 and S2, for the three benchmark
lines 10.1.1, 10.4.1 and 40.2.1, as function of the bino-singlino mass
difference, $\Delta_M$. In each case we first discuss the branching
ratios for the decay $\chi^0_2 \to \chi^0_1 + X$ as function of
$\Delta_M$, which are astonishingly complex since many different
final states contribute for different values of $\Delta_M$. For the
values of $\Delta_M$ where the branching ratios into leptons are large,
we give estimates of the signal cross sections in the multilepton
channels S3. Finally we provide the same information for the cNMSSM
points cNMSSM.1 and cNMSSM.2, where we add estimates of the signal cross
sections in the $2\tau$ channels S4.

\subsection{Benchmark line 10.1.1}

\begin{figure}[ht!]
\begin{center}
\includegraphics[scale=0.5,angle=-90]{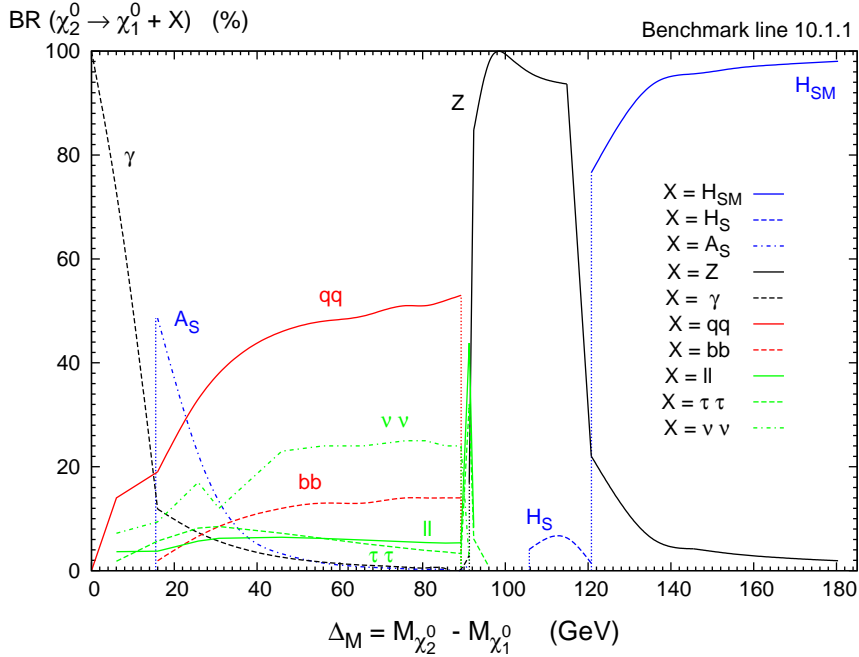}
\end{center}
\caption{Branching fractions into the various states $X$ in the decay
$\chi^0_2 \to \chi^0_1 + X$ as function of $\Delta_M$ for the benchmark
line 10.1.1.}
\label{fig:1}
\end{figure}

The branching ratios for the decay $\chi^0_2 \to \chi^0_1 + X$ as
function of $\Delta_M$ for the benchmark line 10.1.1 (for NMSSM-like
parameters varying as given in Table~\ref{tab:1}) is shown in
Fig.\ref{fig:1}.

As stated before, many different 2-body and 3-body final states ($X$
corresponding to 1-body or 2-body states, respectively) are possible.
For a small bino-singlino mass difference $\Delta_M$, the dominant
$\chi^0_2$ decay mode is the radiatively induced 2-body decay $\chi^0_2
\to \chi^0_1 + \gamma$. However, the $\gamma$ energy would probably be
too small to allow its detection. For larger $\Delta_M$ in the range
$\sim 20-25$~GeV, the 2-body decay into the mostly singlet-like CP-odd
Higgs boson $A_S$ dominates. The possible relevance of $A_S$ production
(and/or $H_S$ production) in neutralino
decays in the NMSSM has already been underlined in
\cite{Cheung:2008rh,Stal:2011cz}. In turn, $A_S$ will decay dominantly
into a pair of $b$-quarks. For $\Delta_M < M_Z$ one finds a plethora of
3-body decays into $q\,\bar{q}$, $b\,\bar{b}$, leptons (electrons and
muons), $\tau^+\,\tau^-$ and neutrinos most of which are mediated by the
$Z$ boson and $A_S$.
For $M_Z < \Delta_M < M_{H_{SM}}$, $\chi^0_2$ decays nearly
exclusively into $\chi^0_2 \to \chi^0_1 + Z$ (with a small branching
fraction into $H_S$), and for $\Delta_M >
M_{H_{SM}}$ nearly exclusively into the SM-like Higgs boson.

Next we apply the event selections and cuts according to the 5 different
signal regions of the ATLAS jets + missing transverse momentum (0
leptons) analysis in \cite{Aad:2011ib} for various values of $\Delta_M$
(see the search channels S1 in Section~3). The same cuts are applied to
the cMSSM point 10.1.1, and subsequently we determine the ratios $R$ of
efficiencies sNMSSM/cMSSM. The results for $R$ are shown in
Fig.~\ref{fig:2}. (Here and in the following Figures for $R$, the error
bars indicate statistical errors only which follow from the fact that we
have simulated about 10000 events per point.)

\begin{figure}[ht!]
\begin{center}
\vspace*{10mm}
\includegraphics[scale=0.5,angle=-90]{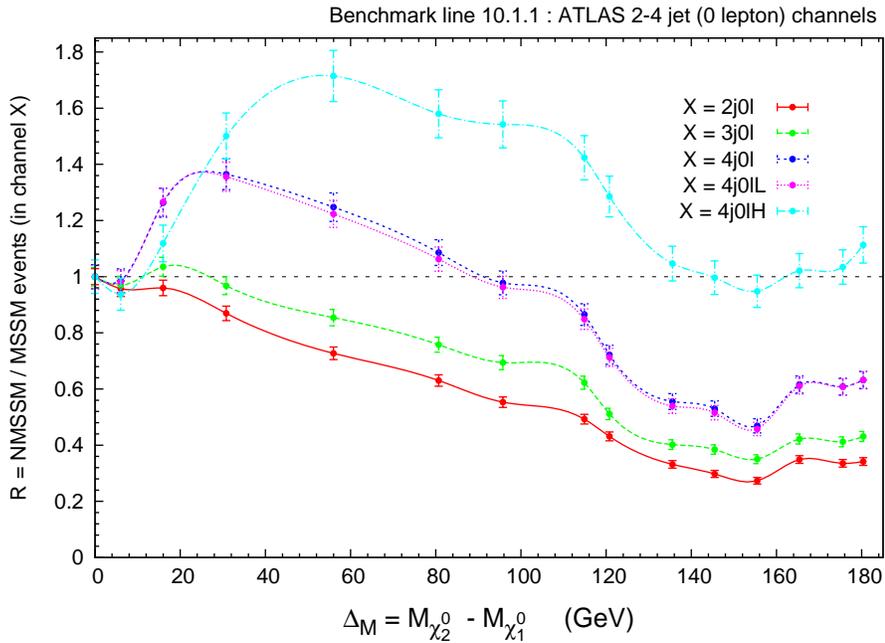}
\end{center}
\caption{The ratios $R$ of efficiencies as function of~$\Delta_M$ in the
sNMSSM w.r.t. the cMSSM point 10.1.1 in 5 different signal regions of
the ATLAS jets + missing transverse momentum analysis.}
\label{fig:2}
\end{figure}

We see that i) $R$ is nearly always larger than 1 for the $4jH$ signal
region; ii) for the $4j$ and $4jL$ signal regions, $R > 1$ for $\Delta_M
\lsim 90$~GeV, but $R < 1$ for $\Delta_M \gsim 100$~GeV, and iii) $R <
1$ everywhere for the $3j$ and $2j$ signal regions. Notably for the
latter, $R$ can drop to $\sim 0.3$ for $\Delta_M \sim 150$~GeV. For
$\Delta_M > 115$~GeV, the dominant $\chi^0_2$ decay is $\chi^0_2 \to
\chi^0_1 + H_{SM}$ and, whereas $H_{SM}$ carries away a considerable
amount of (no longer invisible) transverse energy, its decay products
($b$-jets) contribute to the signatures. Hence, the reduction of $R$ is
dominant for the $3j$ and $2j$ signal regions, which hardly profit from
the additional $b$-jets.

\begin{figure}[ht!]
\begin{center}
\vspace*{10mm}
\includegraphics[scale=0.5,angle=-90]{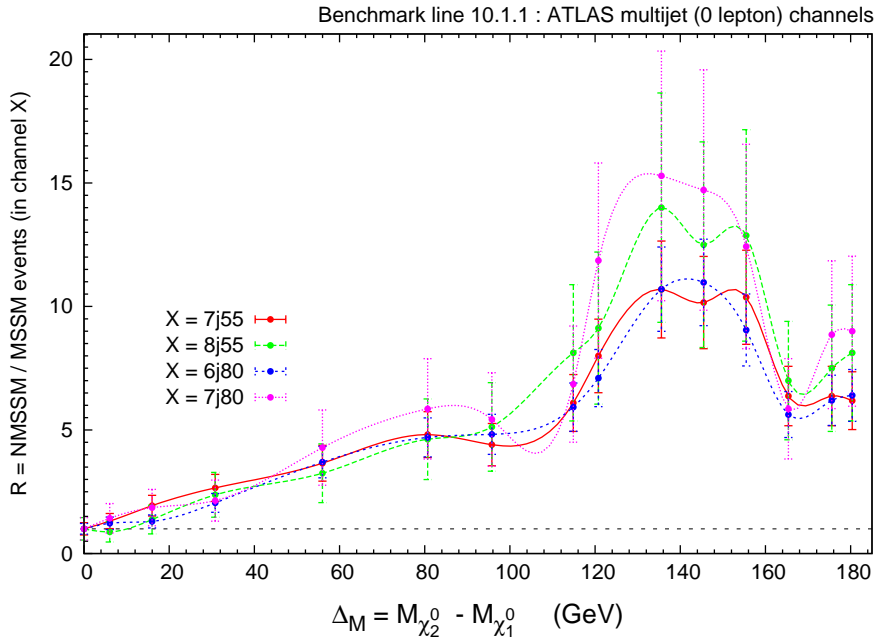}
\end{center}
\caption{The ratios $R$ of efficiencies as function of~$\Delta_M$ in the
sNMSSM w.r.t. the cMSSM point 10.1.1 in 4 different signal regions of
the ATLAS multijet + missing transverse momentum analysis.}
\label{fig:3}
\end{figure}

The impact of the additional jets is also clearly visible in the case of
the ATLAS multijet + missing transverse momentum (0 leptons) analysis in
\cite{Aad:2011qa} (see the search channels S2 in Section~3). We proceed
as above, and study the ratios $R$ of efficiencies in the sNMSSM
w.r.t. the cMSSM in the 4 different signal regions as function of
$\Delta_M$. The results are shown in Fig.~\ref{fig:3}.

Due to the additional jets in the final states and the somewhat weaker
cut on $E_T^\mathrm{miss}$ in \cite{Aad:2011qa}, $R$ can become
considerably larger than 1, notably for $\Delta_M \gsim 115$~GeV where
the $b$-jets from $H_{SM}$ contribute to the signal region.

Given the present absence of clear signals for supersymmetry, an
important question is to which extent the modified efficiencies in the
sNMSSM affect the present lower boundaries in the $m_0- M_{1/2}$ plane.
The answer is not obvious since, depending on the search channel, the
efficiencies in the sNMSSM can be larger or smaller than in the cMSSM.
Hence we have to compare the search channels which are actually relevant
for the most stringent bounds in the $m_0- M_{1/2}$ plane.

The cMSSM point 10.1.1 is part of the benchmark line 10.1 in
\cite{AbdusSalam:2011fc}, which is specified by relatively low values
for $m_0$ and $A_0 = 0$, $\tan\beta=10$. For the same values of $A_0$
and $\tan\beta$, the boundaries in the $m_0- M_{1/2}$ plane implied by
the ATLAS jets + missing transverse momentum analysis in
\cite{Aad:2011ib} can be found, channel by channel, on the web page
\cite{ATLASweb}. One finds that, for these low values of $m_0$, the
constraints are dominated by the $2j$/$3j$ signal regions implying
$M_{1/2} \gsim 470/450$~GeV, respectively. The $2j$/$3j$ signal regions
are precisely those for which the modified efficiencies in the sNMSSM
can be considerably lower (by a factor $\sim 1/3$) than in the cMSSM, as
can be seen in Fig.~\ref{fig:2} for $\Delta_M \sim 150$~GeV.

Hence, the constraints from the absence of excesses in the $2j$/$3j$
signal regions can accommodate, in the sNMSSM with $\Delta_M \sim
150$~GeV, a production cross section which is about 3 times larger than
in the cMSSM, corresponding to a somewhat smaller value for $M_{1/2}$.
Using again the next-to-leading order production cross sections from
Prospino \cite{Beenakker:1996ch,Beenakker:1996ed}, we find that the
latter decrease with increasing $M_{1/2}$ (implying increasing squark
and gluino masses) roughly like $M_{1/2}^{-8.5}$. Thus in the sNMSSM for
$\Delta_M \sim 150$~GeV, the lower bound on $M_{1/2}$ from the absence
of excesses in the $2j/3j$ signal region is lower than in the cMSSM by a
factor $(1/3)^{1/8.5} \sim 0.88$, leading to $M_{1/2} \gsim 415$~GeV
instead of $M_{1/2} \gsim 470$~GeV for $m_0 \sim 125$~GeV, $A_0=0$ and
$\tan\beta=10$.

Of course we must verify whether such a reduced value of $M_{1/2}$ is
consistent with constraints from the other search channels, notably
those in which the efficiencies in the \hbox{sNMSSM} are larger than in the
cMSSM: First, in the remaining search channels in the ATLAS jets +
missing transverse momentum analysis in \cite{Aad:2011ib} the
efficiencies in the sNMSSM with $\Delta_M \sim 150$~GeV are never
enhanced, and $M_{1/2} \gsim 415$~GeV remains consistent with the
corresponding bounds. Also, the lower bounds on $M_{1/2}$ from the ATLAS
multijet + missing transverse momentum analysis in \cite{Aad:2011qa},
for $m_0 \sim 125$~GeV, are so low that the enhanced efficiencies in the
sNMSSM from Fig.~\ref{fig:3} remain consistent with present bounds for
$M_{1/2} \gsim 415$~GeV.

We have all reasons to expect that these conclusions -- a reduced lower
bound on $M_{1/2}$ by a factor $\sim 0.88$ in the sNMSSM with $\Delta_M
\sim 150$~GeV w.r.t. the cMSSM for corresponding values of $m_0$, $A_0$
and $\tan\beta$ -- remain valid in the light of the constraints from CMS
in \cite{Chatrchyan:2011zy}, which are slightly stronger ($M_{1/2} \gsim
540$~GeV, to be replaced by $M_{1/2} \gsim 475$~GeV): Again, the
constraints are dominated by the 2/3 jet analyses (see 
\cite{Khachatryan:2011tk}), for which the efficiencies in the sNMSSM can
be reduced due to less missing transverse momentum, thus being
compatible with a larger production cross section than in the cMSSM. (A
detailed analysis of all available SUSY search channels would go beyond
the scope of the present paper.)

However, it is of interest to verify to which extent these sNMSSM
scenarios would contribute to the CMS analysis of multilepton signatures
in \cite{CMS11-013} (see the search channels S3 in Section~3). The
largest efficiencies in the sNMSSM are found in the region $\Delta_M
\sim 180$~GeV, corresponding to $M_{\chi^0_1} \sim 25$~GeV; the
corresponding signal cross sections (including statistical errors) are
given in Table~\ref{tab:3}.

\begin{table}[h!]
\begin{center}
\begin{tabular}{|c|c|c|c|c|} \hline
Channel:      & MET3 & MET4  & HT3 & HT4 \\\hline
$\sigma$ [fb] & 1.61 $\pm$ 0.09 & 0.29 $\pm$ 0.04 & 1.63 $\pm$ 0.09 &
0.31  $\pm$ 0.04 \\\hline
\end{tabular}
\end{center}
\caption{Signal cross sections for the CMS multilepton search channels
MET3, MET4, HT3 and HT4 (see text) for the sNMSSM with parameters
corresponding to the cMSSM point 10.1.1, and a light singlino-like LSP
with $M_{\chi^0_1} \sim 25$~GeV.}
\label{tab:3}
\end{table}

We see that the values for $\sigma$ are small and would hardly give
visible event rates for luminosities of a few fb$^{-1}$. Hence sNMSSM
scenarios with somewhat lower values of $M_{1/2}$ as compared to cMSSM
scenarios, as allowed by the $2j/3j$ search channels, are not ruled out.
For the corresponding cMSSM point 10.1.1, leptons are expected from the
cascade decay of the wino-like NLSP $\chi^0_2 \to \chi^0_1 +
\bar{l}l/\bar{\tau}\tau$ (see the discussion in
\cite{AbdusSalam:2011fc}). Still, the corresponding signal cross sections
in these channels are smaller by about a factor 1/6 (for MET3/4) or 1/3
(for HT3/4); in the future, such differences can help distinguishing
different scenarios for supersymmetry.

\subsection{Benchmark line 10.4.1}

\begin{figure}[ht!]
\begin{center}
\includegraphics[scale=0.5,angle=-90]{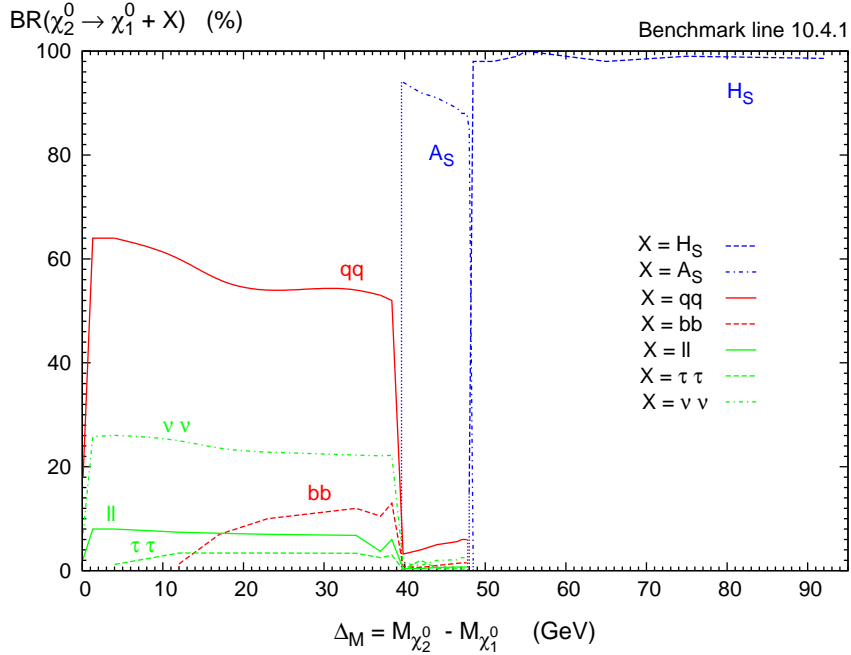}
\end{center}
\caption{Branching fractions into the various states $X$ in the decay
$\chi^0_2 \to \chi^0_1 + X$ as function of $\Delta_M$ for the benchmark
line 10.4.1.} 
\label{fig:4}
\end{figure}

Compared to the sNMSSM benchmark line 10.1.1, the benchmark line 10.4.1
corresponds to a larger value of $\lambda=0.13$ and $m_0=750$~GeV, but a
smaller value for $M_{1/2}=350$~GeV, as given in Table~\ref{tab:1}.  Due
to the larger mixings in the Higgs sector for $\lambda=0.13$, it is more
difficult to satisfy LEP constraints on light Higgs bosons. Hence the
singlino-like neutralino mass has to be larger than $\sim 50$~GeV, or
$\Delta_M \lsim 93$~GeV. The branching fractions into the various states
$X$ in the decay $\chi^0_2 \to \chi^0_1 + X$ are shown in
Fig.~\ref{fig:4}. 

In contrast to Fig.~\ref{fig:1}, the radiatively induced 2-body decay
$\chi^0_2 \to \chi^0_1 + \gamma$ is no longer dominant for a small
bino-singlino mass difference $\Delta_M$. However, for 35~GeV~$\lsim
\Delta_M \lsim 45$~GeV, the 2-body decay into the mostly singlet-like
CP-odd Higgs boson $A_S$ dominates again. Due to the larger value of
$\lambda$ compared to the corresponding analysis of the 10.1.1 line (see
Fig.~\ref{fig:1}), the 2-body decay into the mostly singlet-like $H_S$
dominates for 48~GeV~$\lsim \Delta_M$.

\begin{figure}[!ht]
\begin{center}
\includegraphics[scale=0.45,angle=-90]{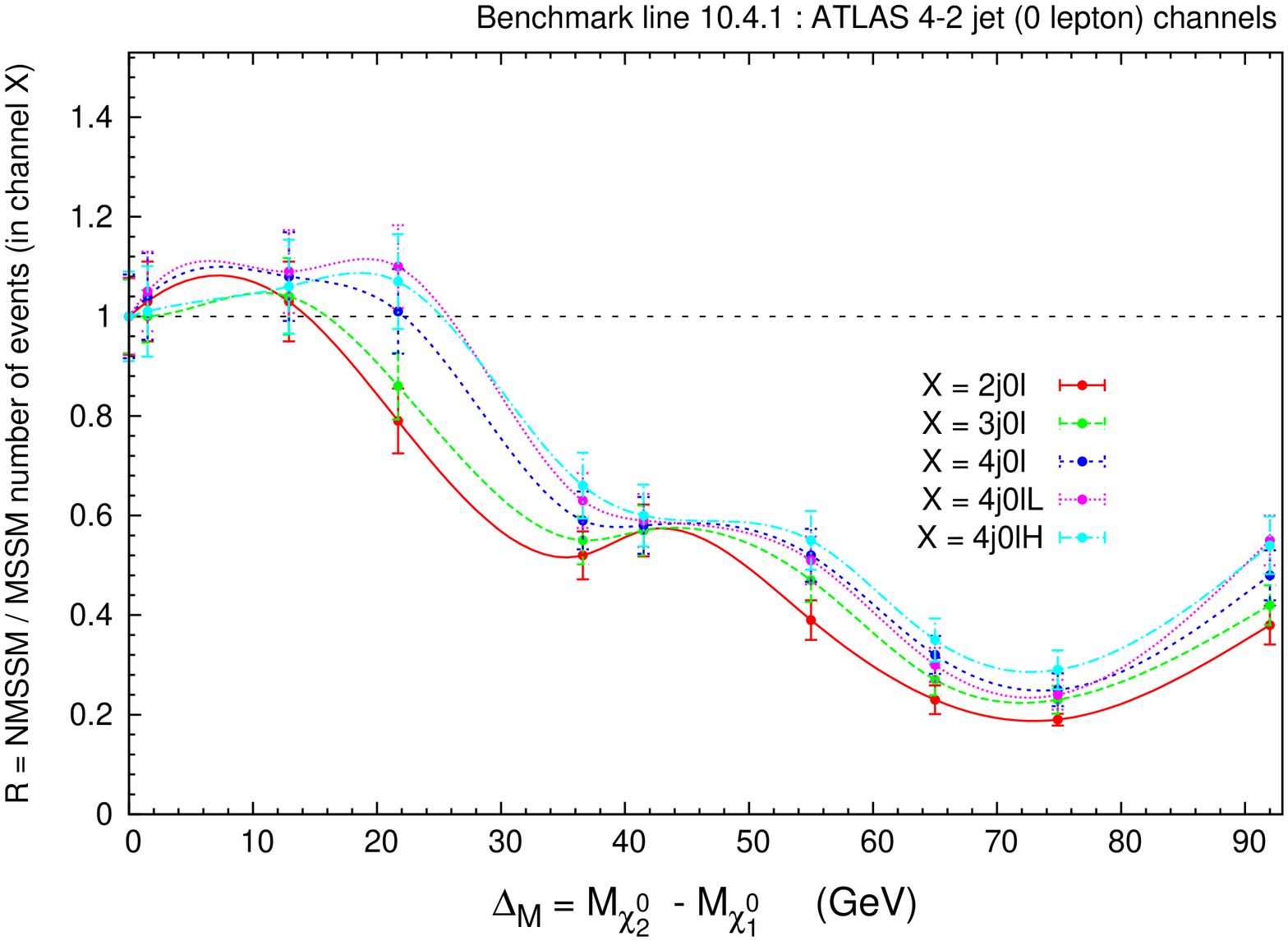}
\vspace*{3mm}
\\
\includegraphics[scale=0.45,angle=-90]{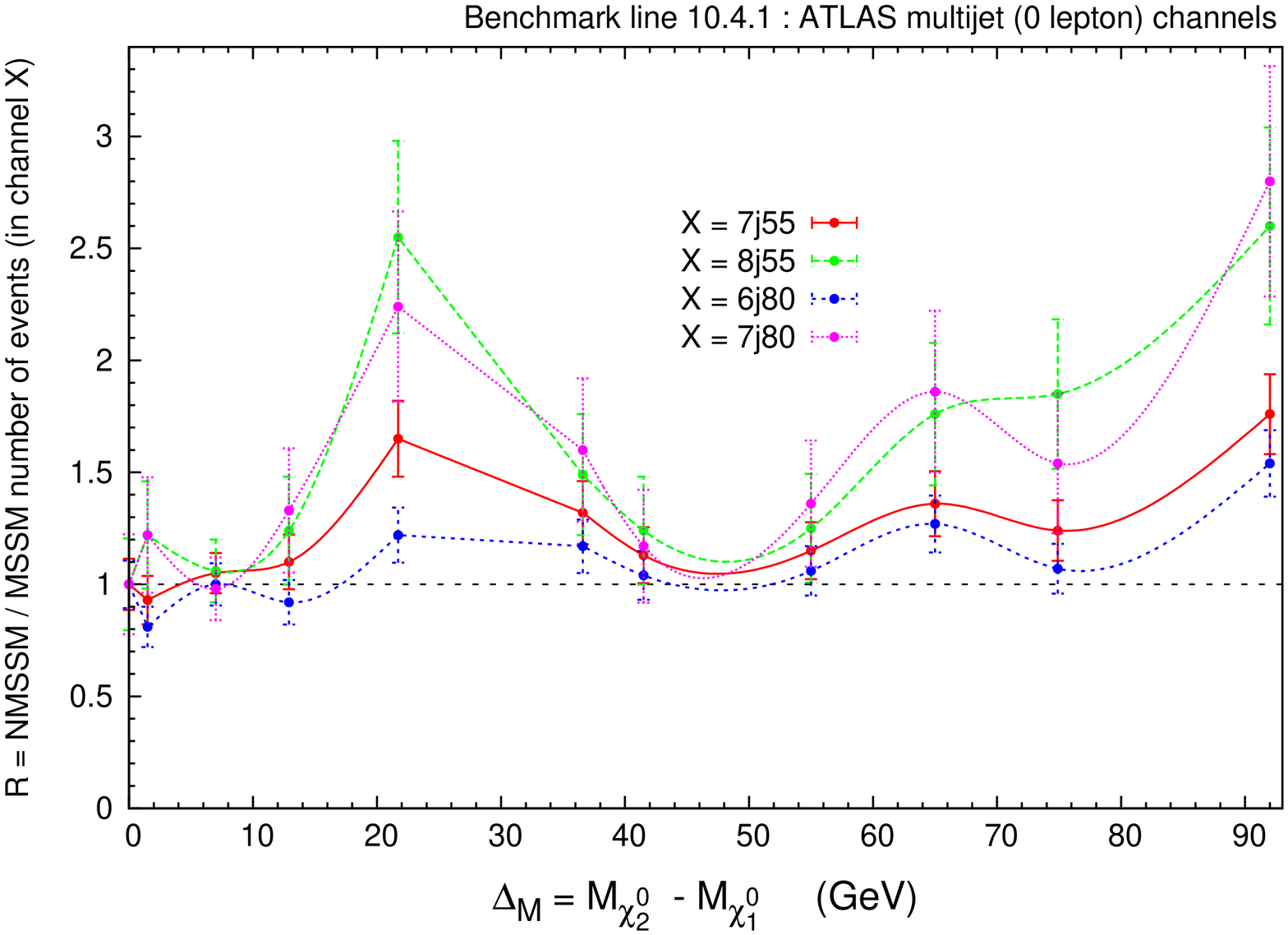}
\end{center}
\vspace*{-3mm}
\caption{The ratios $R$ of efficiencies as function of~$\Delta_M$ in the
sNMSSM w.r.t. the cMSSM point 10.4.1 in 5 different signal regions of
the ATLAS jets + missing transverse momentum analysis (upper panel) and in 4
different signal regions of the ATLAS multijet + missing transverse
momentum analysis (lower panel).}
\label{fig:5}
\end{figure}

Next we consider again the ratios $R$ of efficiencies in the
sNMSSM as a function of $\Delta_M$ w.r.t. the cMSSM, for the ATLAS jets
+ missing transverse momentum analysis in \cite{Aad:2011ib} (search
channels S1) and the ATLAS multijet + missing transverse momentum
analysis in \cite{Aad:2011qa} (search channels S2). The results are
shown in Figs.~\ref{fig:5}.

In the case of the ATLAS jets + missing transverse momentum analysis -
the upper panel of Figs.~\ref{fig:5} - we see that now $R$ can
decrease to $\sim 0.3$ for all search channels for $\Delta_M \sim
75$~GeV ($M_{\chi^0_1} \sim 70$~GeV), becoming as small as $R$ $\sim
0.2$ for the most relevant $2j/3j$ channels. The increase of $R$ for
the ATLAS multijet + missing transverse momentum analysis shown in the
lower panel of Figs.~\ref{fig:5} is less pronounced than in
Fig.~\ref{fig:3}. As explained in \cite{AbdusSalam:2011fc}, the reason
is that here the gluino is lighter than the squarks and its dominant
3-body decays yield higher fractions of final states with more hadronic
jets, already in the cMSSM. Hence the \emph{relative} increase of
multijet efficiencies in the sNMSSM is smaller than for the previous
line 10.1.1.

As before, we can estimate to what extent the reduced value of $R \sim
0.2$ for the most relevant $2j/3j$ channels alleviates the lower bound
on $M_{1/2}$: For $m_0=750$~GeV, the production cross sections for
squarks/gluinos decrease roughly like $M_{1/2}^{-6.3}$. Thus in the
sNMSSM for $\Delta_M \sim 75$~GeV, the lower bound on $M_{1/2}$ from the
absence of excesses in the $2j/3j$ signal region is lower than in the cMSSM
by a factor $(1/5)^{1/6.3} \sim 0.75$. Again this conclusion is not
affected by the larger efficiencies in the less sensitive multijet
channels.

The largest efficiencies in the sNMSSM in the multilepton channels
MET3/HT3 analysed by CMS in \cite{CMS11-013} are found in the region $\Delta_M
\sim 45$~GeV corresponding to $M_{\chi^0_1} \sim 100$~GeV originating
from tau leptons arising from $A_S$ decays (see Fig.~\ref{fig:4}); the
associated signal cross sections are given in Table~\ref{tab:4}. (The
signal cross sections in the MET4/HT4 channels, requiring 4 leptons
passing the cuts, are even smaller than for the 10.1.1 point shown in
Table~\ref{tab:3} and of the order of our statistical errors.)

\begin{table}[h!]
\begin{center}
\begin{tabular}{|c|c|c|c|c|} \hline
Channel:      & MET3 & MET4  & HT3 & HT4 \\\hline
$\sigma$ [fb] & $2.6\pm 0.2$ & $\lsim 0.1$ & $1.5\pm 0.1$ & $\lsim
0.1$\\\hline
\end{tabular}
\end{center}
\caption{Signal cross sections for the CMS multilepton search channels
for the sNMSSM with parameters corresponding to the cMSSM point 10.4.1,
and a singlino-like LSP with $M_{\chi^0_1} \sim 100$~GeV.}
\label{tab:4}
\end{table}

We observed that events originating from squark/gluino production
contribute simultaneously to the MET and HT search channels, and would
give practically identical signal cross sections. However, events
originating from neutralino/chargino/slepton production only contribute
to the MET search channels, since they lead to less jets which would be
required for the HT cuts; this explains the slightly larger signal cross
section in MET3 compared to HT3. In the
cMSSM, the signal cross sections are smaller by about a factor 1/3, but
even in the NMSSM
the event rates are obviously small for luminosities of a few~fb$^{-1}$.

\subsection{Benchmark line 40.2.1}

\begin{figure}[ht!]
\begin{center}
\includegraphics[scale=0.5,angle=-90]{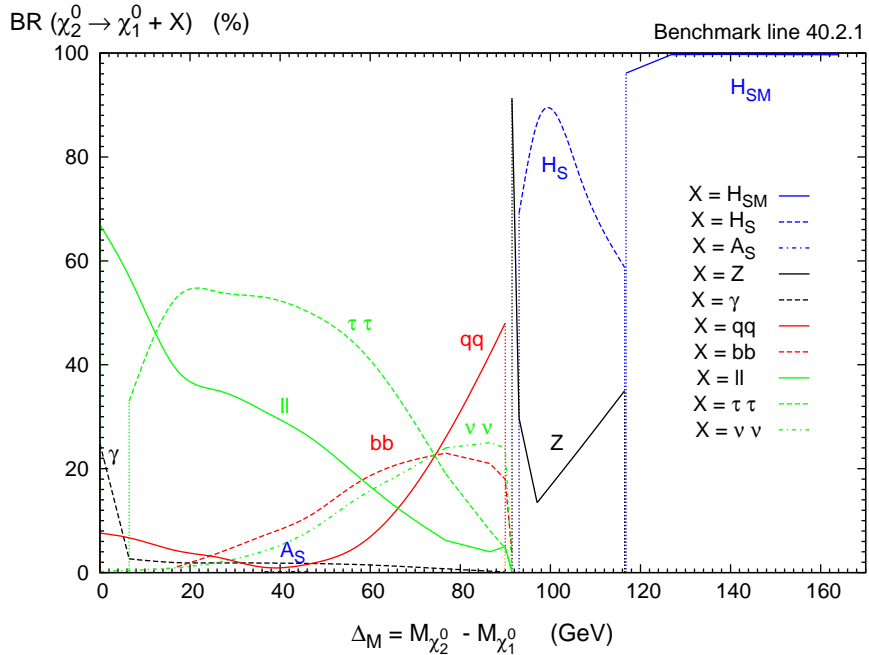}
\end{center}
\caption{Branching fractions into the various states $X$ in the decay
$\chi^0_2 \to \chi^0_1 + X$ as function of $\Delta_M$ for the benchmark
line 40.2.1.}
\label{fig:6}
\end{figure}

Now we study a region in parameter space with larger values of
$\tan\beta=40$ and $A_0=-500$~GeV, as defined in Table~\ref{tab:1}. The
branching fractions into the various states $X$ in the decay $\chi^0_2
\to \chi^0_1 + X$ are shown in Fig.~\ref{fig:6}.
In this case, the 3-body decays of $\chi^0_2$ into leptons, taus and
quark pairs dominate for $\Delta_M \lsim 90$~GeV. For 90~GeV~$\lsim
\Delta_M \lsim 115$~GeV, 2-body decays into $H_S$ (and $Z$)
dominate whereas, as before, 2-body decays into $H_{SM}$ dominate for
$\Delta_M \gsim 115$~GeV.

\begin{figure}[!ht]
\vspace*{5mm}
\begin{center}
\includegraphics[scale=0.45,angle=-90]{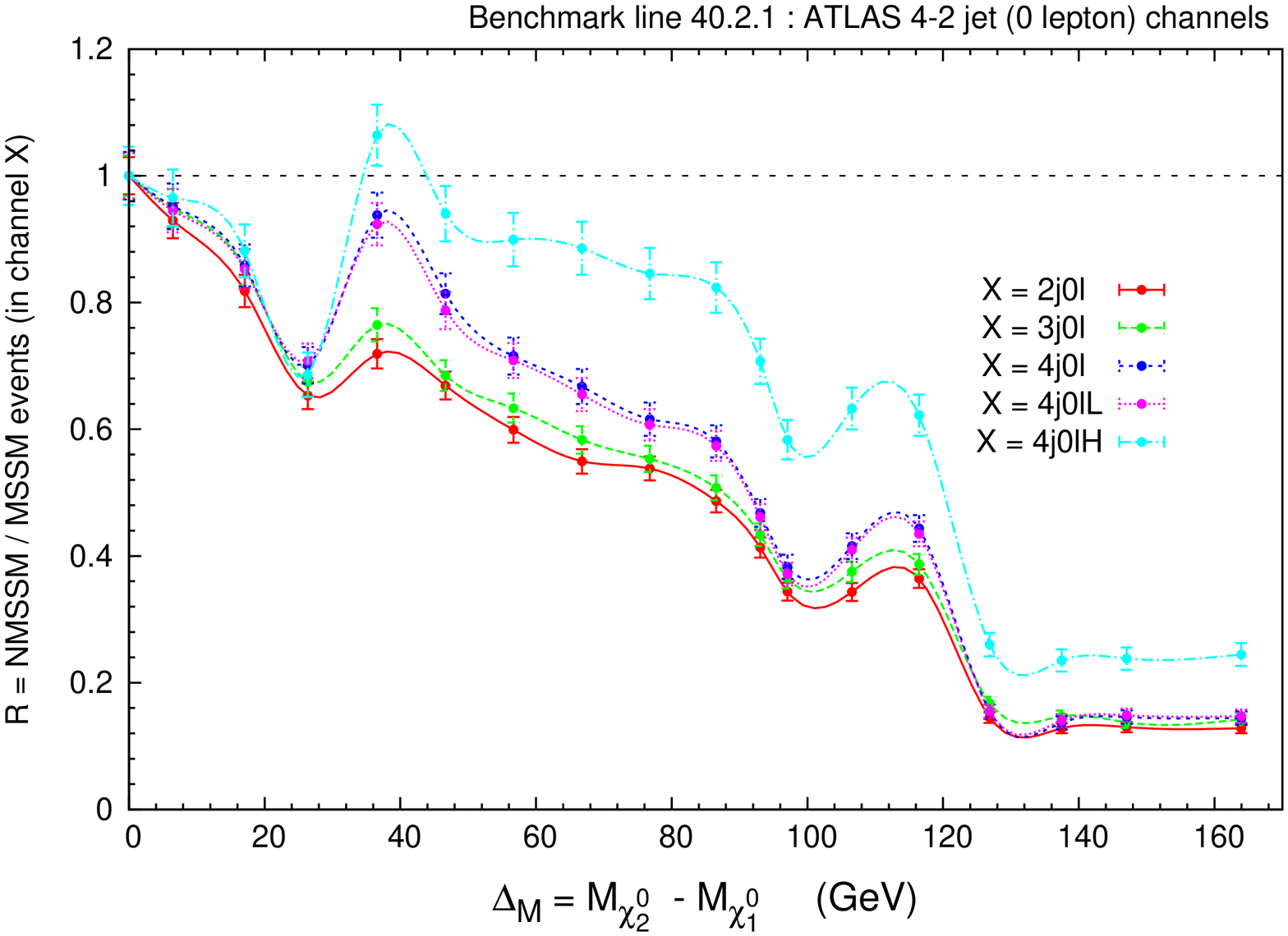}
\vspace*{5mm}
\\
\includegraphics[scale=0.45,angle=-90]{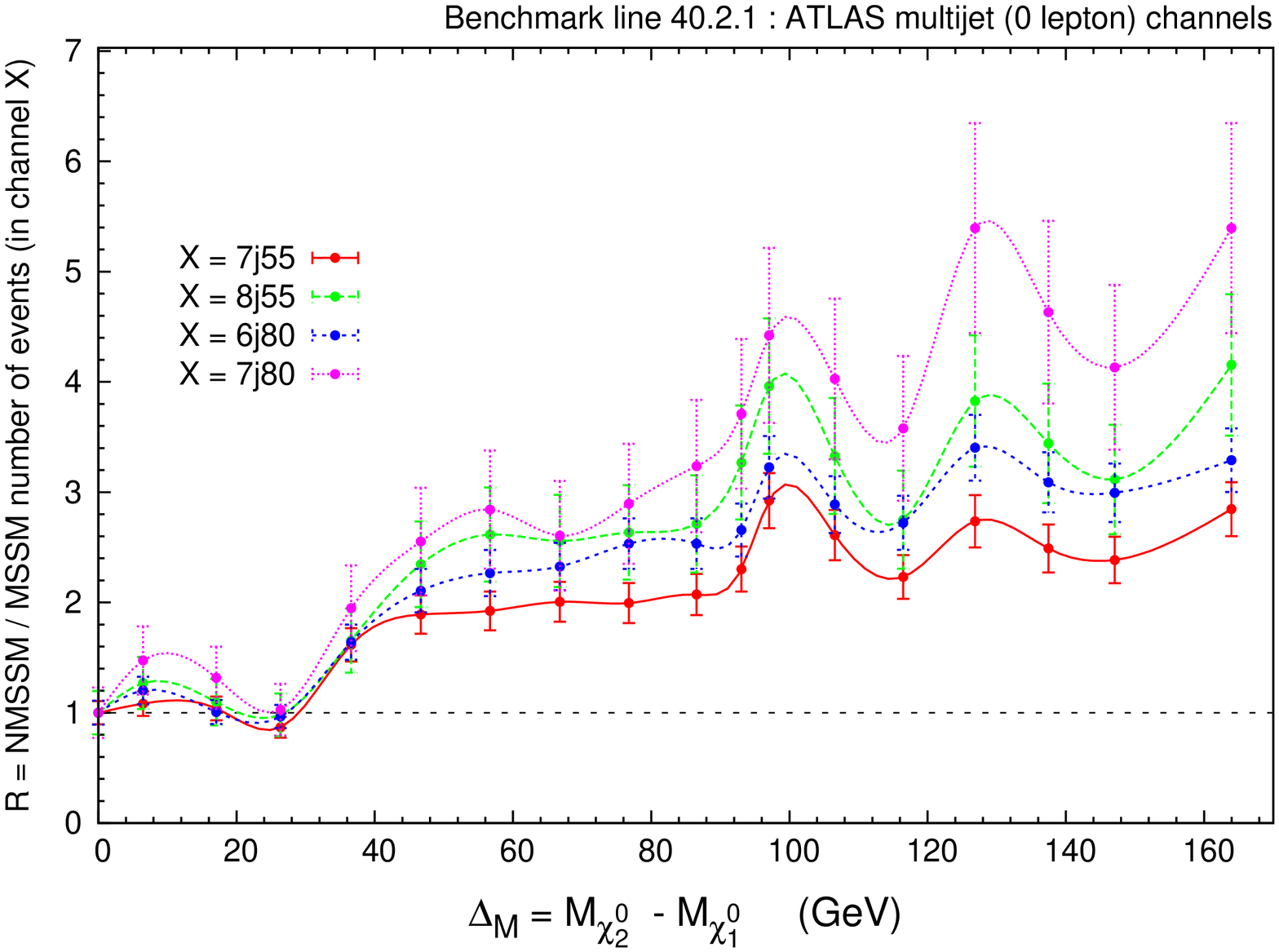}
\end{center}
\vspace*{-3mm}
\caption{The ratios $R$ of efficiencies as function of~$\Delta_M$ in the
sNMSSM w.r.t. the cMSSM point 40.2.1 in 5 different signal regions of
the ATLAS jets + missing transverse momentum analysis (upper panel) and in 4
different signal regions of the ATLAS multijet + missing transverse
momentum analysis (lower panel).}
\label{fig:7}
\end{figure}

For the search channels S1 and S2, the results for the ratios $R$ of
efficiencies in the sNMSSM as function of $\Delta_M$ w.r.t. the cMSSM
are shown in Figs.~\ref{fig:7}.

Here the decrease of $R$ for the ATLAS jets + missing transverse momentum
analysis is even stronger than in the previous cases; for the most
relevant $2j$ and $3j$ channels, $R$ drops below 0.15 for $\Delta_M \gsim
130$~GeV ($M_{\chi^0_1} \lsim 60$~GeV). For the ATLAS multijet + missing
transverse momentum analysis $R$ increases up to $\sim 5$ in this region
of $\Delta_M$, which is again less pronounced than in the 10.1.1
analysis of Fig.~\ref{fig:3}.  The reason is that already in the cMSSM
the gluino decays dominantly into a stop+top pair
\cite{AbdusSalam:2011fc}, yielding again (but for a different reason)
higher fractions of final states with more hadronic jets making the
\emph{relative} increase of multijet efficiencies in the sNMSSM smaller
than for the point 10.1.1.

Concerning the multilepton channels analysed by CMS, the largest
efficiencies in the sNMSSM are found in the region $\Delta_M \sim
145$~GeV corresponding to $M_{\chi^0_1} \sim 40$~GeV; the corresponding
signal cross sections are given in Table~\ref{tab:5}.

\begin{table}[ht!]
\begin{center}
\begin{tabular}{|c|c|c|c|c|} \hline
Channel:      & MET3 & MET4  & HT3 & HT4 \\\hline
$\sigma$ [fb] & 2.72 $\pm$ 0.03 & 0.11 $\pm$ 0.01 & 2.72 $\pm$ 0.03 &
0.11$\pm$ 0.01 \\\hline
\end{tabular}
\end{center}
\caption{Signal cross sections for the CMS multilepton search channels
for the sNMSSM with parameters corresponding to the cMSSM point 40.2.1,
and a light singlino-like LSP with $M_{\chi^0_1} \sim 40$~GeV.}
\label{tab:5}
\end{table}

For the cMSSM point 40.2.1, these signal cross sections are smaller by
a factor less than 1/10. The origin of the larger multilepton signal
cross section in the sNMSSM is the 3-body decay of $\chi^0_2$ into
$\chi^0_1$ plus leptons, as shown in Fig.~\ref{fig:6} (see also
\cite{arXiv:0811.0011,Barger:2010aq}), with leptons sufficiently
energetic to survive the cuts in the CMS analysis.

Since in the sNMSSM the reduction of $R$ down to 0.15 in the most
relevant $2j$/$3j$ channels is stronger than before, the sNMSSM in this
region of parameter space could be compatible, in the absence of
signals, with sparticle production cross sections about 7~times larger
than in the cMSSM. Since here (for $m_0 = 550$~GeV, $A_0=-500$~GeV,
$\tan\beta=40$) the squark/gluino production cross section decreases
$\sim M_{1/2}^{-7.5}$, for $\Delta_M \gsim 130$~GeV the cNMSSM is
compatible with values of $M_{1/2}$ which are smaller than in the 
cMSSM by about 0.75. 
(Again this conclusion is not affected by the enhancement
of $R$ in the less sensitive multijet or multilepton channels.) These
potential attenuations of lower bounds on $M_{1/2}$ are not dramatic,
but neither completely negligible.

\subsection{The cNMSSM}

The parameters and some sparticle and Higgs masses for two cNMSSM
benchmark points, as well as nearby cMSSM points in the stau
co-annihilation region, have been given in Table~\ref{tab:2}. In this
case the $\tilde{\tau}_1$ is the NLSP, decaying as $\tilde{\tau}_1 \to
\tau + \chi^0_1$, where $\chi^0_1$ is mostly singlino-like in the
cNMSSM, but mostly bino-like for the cMSSM points. Due to the small
$\tilde{\tau}_1 - \chi^0_1$ mass difference, these $\tau$ leptons are
however quite soft. Harder $\tau$ leptons appear in the sparticle decays
into $\tilde{\tau}_1$. One such sparticle decay appears in every
sparticle decay cascade in the cNMSSM, but only occasionally in the
cMSSM in the stau co-annihilation region.

First we compare, as before, the ratios $R$ of efficiencies in the cNMSSM
w.r.t. the cMSSM in 5 different signal regions of the ATLAS jets +
missing transverse momentum analysis (search channels S1); the results
are given in Table~\ref{tab:6}.

\begin{table}[ht!]
\begin{center}
\begin{tabular}{|c|c|c|c|c|c|} \hline
Channel:      & $2j$ & $3j$  & $4j$ & $4jL$ & $4jH$ \\\hline
$R_1$ & $0.63\pm 0.04$ & $0.73\pm 0.05$ & $0.86\pm 0.06$ 
& $0.86\pm 0.06$ & $0.95\pm 0.08$ \\\hline
$R_2$ & $0.65\pm 0.03$ & $0.75\pm 0.03$ & $0.95\pm 0.06$
& $0.96\pm 0.06$ & $1.1\pm 0.1$ \\\hline
\end{tabular}
\end{center}
\caption{Ratios $R_1 =$(cNMSSM.1/cMSSM.1) and $R_2 =$(cNMSSM.2/cMSSM.2) 
of efficiencies for two points of the cNMSSM
w.r.t. the cMSSM in 5 different signal regions of the ATLAS jets +
missing transverse momentum analysis.}
\label{tab:6}
\end{table}

With $R_{1.2} \sim 0.6-1.1$, the differences between the cNMSSM and the cMSSM
are not spectacular in these search channels. In the case of the
multijets + missing transverse momentum analysis (S2), efficiencies are
so small that a comparison is not meaningful. In Table~\ref{tab:7} we
give the signal cross sections for the multilepton search channels
analysed by CMS (S3).

\begin{table}[ht!]
\begin{center}
\begin{tabular}{|c|c|c|c|c|} \hline
Channel:      & MET3 & MET4  & HT3 & HT4  \\\hline
$\sigma$ [fb] (cNMSSM.1) & $4.4\pm 0.2$ & $0.7\pm 0.1$ & $2.7\pm 0.1$
 & $0.41\pm 0.05$ \\\hline
$\sigma$ [fb] (cNMSSM.2) & $1.63\pm 0.06$ & $0.36\pm 0.03$ & $0.94\pm
 0.04$ & $0.18\pm 0.02$ \\\hline
\end{tabular}
\end{center}
\caption{Signal cross sections for the CMS multilepton search channels
for two points of the cNMSSM.}
\label{tab:7}
\end{table}

In spite of the smaller total sparticle production cross sections in the
cNMSSM compared to the sNMSSM benchmark lines, the signal cross sections
in the multilepton search channels are at least of the same order,
notably for the MET channels, where no hard jets are required: Here
neutralino/chargino/slepton/stau production processes contribute, in
contrast to the HT channels which require hard jets originating from
squark/gluino production. (For the corresponding cMSSM points cMSSM.1
and cMSSM.2, these signal cross sections are smaller by about a factor
1/20 and not shown here.) In \cite{CMS11-013}, 2.1~fb$^{-1}$ of
luminosity have been analysed by CMS in the multilepton channels. No
significant excesses are expected within the cNMSSM at present, but
these search channels can become sensitive to the cNMSSM in the future.

Due to the large number of $\tau$ leptons in the final states, it
becomes interesting to study the three signal regions of the CMS
$2\,\tau$ search channels (S4, see Section~3) for the cNMSSM; the
corresponding signal cross sections are shown in Table~\ref{tab:8}.
\begin{table}[ht!]
\begin{center}
\begin{tabular}{|c|c|c|c|} \hline
Channel:      &$e/\mu\,\tau_h$~high~$E_T^{miss}$ &
$e/\mu\,\tau_h$~high~$H_T$&
$\tau_h\,\tau_h$ \\\hline
$\sigma$ [fb] (cNMSSM.1) & $2.2\pm 0.1$ & $2.4\pm 0.1$ &
$0.65\pm 0.05$\\\hline
$\sigma$ [fb] (cNMSSM.2) & $0.77\pm 0.03$ & $0.81\pm 0.03$ 
& $0.19\pm 0.01$\\\hline
\end{tabular}
\end{center}
\caption{Signal cross sections for the CMS $2\,\tau$ search channels
for two cNMSSM points.}
\label{tab:8}
\end{table}

(Again, for the corresponding cMSSM points cMSSM.1 and cMSSM.2 these
signal cross sections are smaller by about a factor 1/20.) In
\cite{Ellwanger:2010es} it was estimated that only the LHC at 14~TeV
c.m.~energy would become sensitive to the cNMSSM. However, combining
multilepton and $2\,\tau$ search channels and increasing the integrated
luminosity to $\sim 20$~fb$^{-1}$, already the LHC with 7~TeV
c.m.~energy could become sensitive to the low $M_{1/2}$ regime of the
cNMSSM in the future.

\section{Conclusions and Outlook}

In the NMSSM with a singlino-like LSP, it is easy to estimate
qualitatively how the signatures for supersymmetry are modified with
respect to the MSSM: The additional cascade $\chi^0_2 \to \chi^0_1 + X$
will reduce the missing energy, but will provide additional jets or
leptons. In the present paper we studied this issue quantitatively for
several benchmark points (lines in the sNMSSM), considering several
supersymmetry search channels. For the considered benchmark lines, we
found that the efficiencies can drop by a factor $\sim 1/3$ to $\sim
1/7$ with respect to the MSSM in the most relevant 2/3~jet + missing
energy search channels. This can reduce the present lower bounds on
$M_{1/2}$ by a factor $\sim 0.9 - 0.75$ for parameter regions of the
sNMSSM corresponding to large bino--singlino mass differences. The
corresponding increase of efficiencies in multijet or multilepton search
channels is not strong enough to affect this conclusion.

In addition, we studied the cNMSSM, in which the $\tilde{\tau}_1$ is the
NLSP and the $\tilde{\tau}_1$--singlino mass difference is small (in
order to comply with the WMAP bounds on the dark matter relic density),
for the lowest possible values of $M_{1/2}$. cMSSM points with similar
values of $m_0$ and $M_{1/2}$ are not excluded by present searches.
Since the efficiencies in the most relevant 2/3~jet + missing energy
channels in the cNMSSM turn out to be similar, present searches are not
sensitive to the cNMSSM either. However, in the future the signal cross
sections in the multilepton and
$2\,\tau$ search channels could give hints for the cNMSSM already at the
LHC with 7~TeV c.m.~energy for larger integrated luminosities.

Clearly, in a first study of this kind we could only ``scratch the tip
of the iceberg'': First, many more search channels (different
combinations of and cuts on missing transverse energy, jets and leptons)
are studied by the ATLAS and CMS collaborations. Second, we did not
cover even the sNMSSM parameter space completely; we confined ourselves
to regions similar to cMSSM benchmark points with moderate (but not
excluded) values of both $m_0$ and $M_{1/2}$. As in the MSSM, the
parameter space of the general NMSSM is much larger, and studies similar
to those within the general MSSM \cite{Sekmen:2011cz,Arbey:2011un,
Papucci:2011wy, Allanach:2011qr} could reveal more regions in the NMSSM
to which the SUSY search channels are less sensitive than in the MSSM.
Third, more refined studies of efficiencies in various search channels
as function of the final states in the additional bino~$\to$~singlino
cascade can help to clarify under which circumstances the NMSSM can be
distinguished from the MSSM independently from the Higgs sector. Hence
the present study can and should be extended in many different ways.

\section*{Acknowledgements}

U.~E. acknowledges support from the French ANR LFV-CPV-LHC.


\begin{thebibliography}{99}

\bibitem{Ellwanger:2009dp}
  U.~Ellwanger, C.~Hugonie and A.~M.~Teixeira,
  Phys.\ Rept.\  {\bf 496} (2010) 1\newline
  [arXiv:0910.1785 [hep-ph]].

\bibitem{Aad:2011ib}
  G.~Aad {\it et al.}  [ATLAS Collaboration],
  ``Search for squarks and gluinos using final states with jets and
  missing transverse momentum with the ATLAS detector in sqrt(s) = 7 TeV
  proton-proton collisions,''
  arXiv:1109.6572 [hep-ex].

\bibitem{ATLAS:2011ad}
  G.~Aad {\it et al.}  [ATLAS Collaboration],
  Phys.\ Rev.\ D {\bf 85} (2012) 012006
  [arXiv:1109.6606 [hep-ex]].

\bibitem{Aad:2011qa}
  G.~Aad {\it et al.}  [Atlas Collaboration],
  JHEP {\bf 1111} (2011) 099
  [arXiv:1110.2299 [hep-ex]].

\bibitem{1110:6189} 
  G.~Aad {\it et al.} [ATLAS Collaboration],
  ``Searches for supersymmetry with the ATLAS detector using final
  states with two leptons and missing transverse momentum in sqrt{s} = 7
  TeV proton-proton collisions,'' 1110:6189 [hep-ex].

\bibitem{1111:4116}
  G.~Aad {\it et al.} [ATLAS Collaboration],
  ``Search for Diphoton Events with Large Missing Transverse Momentum in
  1 fb$^-1$ of 7 TeV Proton-Proton Collision Data with the ATLAS
  Detector,'' 1111:4116 [hep-ex].

\bibitem{:2011cw}
  [ATLAS Collaboration],
  ``Search for scalar bottom pair production with the ATLAS detector in
  pp Collisions at sqrt{s} = 7 TeV,'' arXiv:1112.3832 [hep-ex].


\bibitem{Khachatryan:2011tk}
  V.~Khachatryan {\it et al.}  [CMS Collaboration],
  Phys.\ Lett.\  B {\bf 698} (2011) 196\newline
  [arXiv:1101.1628 [hep-ex]].

\bibitem{Chatrchyan:2011zy}
  S.~Chatrchyan {\it et al.}  [CMS Collaboration],
  Phys.\ Rev.\ Lett.\  {\bf 107} (2011) 221804
  [arXiv:1109.2352 [hep-ex]].

\bibitem{CMS11-004} CMS Collaboration,
  ``Search for supersymmetry in all-hadronic events with missing
  energy,'' CMS-PAS-SUS-11-004.
  
\bibitem{CMS11-005} CMS Collaboration,
  ``Search for supersymmetry in all-hadronic events with MT2,''
  CMS-PAS-SUS-11-005.

\bibitem{CMS11-007} CMS Collaboration,
  ``Search for supersymmetry in all-hadronic events with tau
    leptons'', CMS-PAS-SUS-11-007.

\bibitem{CMS11-008} CMS Collaboration,
  ``Search for supersymmetry with the razor variables at CMS'',
    CMS-PAS-SUS-11-008.

\bibitem{CMS11-009} CMS Collaboration,
  ``Search for Supersymmetry in Events with Photons, Jets and Missing
  Energy,'' CMS-PAS-SUS-11-009.

\bibitem{CMS11-010} CMS Collaboration,
  ``Search for new physics with same-sign isolated dilepton
    events with jets and missing energy'', CMS-PAS-SUS-11-010.

\bibitem{CMS11-011} CMS Collaboration,
  ``Search for new physics in events with opposite-sign
    dileptons and missing transverse energy'', CMS-PAS-SUS-11-011.

\bibitem{CMS11-013} CMS Collaboration,
  ``Multileptonic SUSY searches,'' CMS-PAS-SUS-11-013.

\bibitem{CMS11-015} CMS Collaboration,
  ``Search for new physics with single-leptons at the LHC,''
  CMS-PAS-SUS-11-015.

\bibitem{Sekmen:2011cz}
  S.~Sekmen, S.~Kraml, J.~Lykken, F.~Moortgat, S.~Padhi, L.~Pape,
  M.~Pierini and H.~B.~Prosper {\it et al.}, ``Interpreting LHC SUSY
  searches in the phenomenological MSSM,'' arXiv:1109.5119 [hep-ph].

\bibitem{Arbey:2011un}
  A.~Arbey, M.~Battaglia and F.~Mahmoudi,
  Eur.\ Phys.\ J.\  C {\bf 72} (2012) 1847
  [arXiv:1110.3726 [hep-ph]].

\bibitem{Papucci:2011wy}
  M.~Papucci, J.~T.~Ruderman and A.~Weiler,
  ``Natural SUSY Endures,''
  arXiv:1110.6926 [hep-ph].
  
\bibitem{Allanach:2011qr}
  B.~C.~Allanach, T.~J.~Khoo and K.~Sakurai,
  JHEP {\bf 1111} (2011) 132
  [arXiv:1110.1119 [hep-ph]].

\bibitem{Ellwanger:2011sk}
  U.~Ellwanger,
  Eur.\ Phys.\ J.\  C {\bf 71} (2011) 1782
  [arXiv:1108.0157 [hep-ph]].

\bibitem{AbdusSalam:2011fc}
  S.~S.~AbdusSalam {\it et al.},
  Eur.\ Phys.\ J.\  C {\bf 71} (2011) 1835
  [arXiv:1109.3859 [hep-ph]].

\bibitem{ATLASweb}
  {\sf
   https://atlas.web.cern.ch/Atlas/GROUPS/PHYSICS/PAPERS/SUSY-2011-07/}.

\bibitem{arXiv:0803.0253}
  A.~Djouadi, U.~Ellwanger and A.~M.~Teixeira,
  Phys.\ Rev.\ Lett.\ \ {\bf 101} (2008) 101802
  [arXiv:0803.0253 [hep-ph]].  
  
\bibitem{arXiv:0811.2699}
  A.~Djouadi, U.~Ellwanger and A.~M.~Teixeira,
  JHEP\ {\bf 0904} (2009) 031
  [arXiv:0811.2699 [hep-ph]].

\bibitem{Ellwanger:2010es}
  U.~Ellwanger, A.~Florent and D.~Zerwas,
  JHEP {\bf 1101} (2011) 103
  [arXiv:1011.0931 [hep-ph]].

\bibitem{Ellwanger:2006rn}
  U.~Ellwanger and C.~Hugonie,
  Comput.\ Phys.\ Commun.\  {\bf 177} (2007) 399
  [arXiv:hep-ph/0612134].

\bibitem{Ellwanger:2004xm}
  U.~Ellwanger, J.~F.~Gunion and C.~Hugonie,
  JHEP {\bf 0502} (2005) 066
  [arXiv:hep-ph/0406215].
  
\bibitem{Ellwanger:2005dv}
  U.~Ellwanger and C.~Hugonie,
  Comput.\ Phys.\ Commun.\  {\bf 175} (2006) 290
  [arXiv:hep-ph/0508022].

\bibitem{Beenakker:1996ch}
  W.~Beenakker, R.~Hopker, M.~Spira and P.~M.~Zerwas,
  Nucl.\ Phys.\  B {\bf 492} (1997) 51
  [arXiv:hep-ph/9610490].
  
\bibitem{Beenakker:1996ed}
  W.~Beenakker, R.~Hopker and M.~Spira,
  ``PROSPINO: A program for the PROduction of Supersymmetric Particles In
  Next-to-leading Order QCD,''
  arXiv:hep-ph/9611232, for updates see
  {\sf http://www.thphys.uni-heidelberg.de/{$\sim$}plehn/prospino/}

\bibitem{Beenakker:1999xh}
  W.~Beenakker, M.~Klasen, M.~Kramer, T.~Plehn, M.~Spira and P.~M.~Zerwas,
  Phys.\ Rev.\ Lett.\  {\bf 83} (1999) 3780
  [Erratum-ibid.\  {\bf 100} (2008) 029901]
  [arXiv:hep-ph/9906298].

\bibitem{Alwall:2011uj}
  J.~Alwall, M.~Herquet, F.~Maltoni, O.~Mattelaer and T.~Stelzer,
  JHEP {\bf 1106} (2011) 128
  [arXiv:1106.0522 [hep-ph]].

\bibitem{Sjostrand:2006za}
  T.~Sjostrand, S.~Mrenna and P.~Z.~Skands,
  JHEP {\bf 0605} (2006) 026
  [hep-ph/0603175].

\bibitem{Alwall:2008qv}
  J.~Alwall, S.~de Visscher and F.~Maltoni,
  JHEP {\bf 0902} (2009) 017
  [arXiv:0810.5350 [hep-ph]].

\bibitem{Das:2011dg}
  D.~Das, U.~Ellwanger and A.~M.~Teixeira,
  Comput.\ Phys.\ Commun.\  {\bf 183} (2012) 774
  [arXiv:1106.5633 [hep-ph]].

\bibitem{Muhlleitner:2003vg}
  M.~Muhlleitner, A.~Djouadi and Y.~Mambrini,
  Comput.\ Phys.\ Commun.\  {\bf 168} (2005) 46
  [arXiv:hep-ph/0311167].

\bibitem{Ovyn:2009tx}
  S.~Ovyn, X.~Rouby and V.~Lemaitre,
  ``DELPHES, a framework for fast simulation of a generic collider
  experiment,'' arXiv:0903.2225 [hep-ph].

\bibitem{Cacciari:2005hq}
  M.~Cacciari and G.~P.~Salam,
  Phys.\ Lett.\  B {\bf 641} (2006) 57
  [arXiv:hep-ph/0512210].

\bibitem{Cheung:2008rh}
  K.~Cheung and T.~-J.~Hou,
  Phys.\ Lett.\ B {\bf 674} (2009) 54
  [arXiv:0809.1122 [hep-ph]].

\bibitem{Stal:2011cz}
  O.~Stal and G.~Weiglein,
  JHEP {\bf 1201} (2012) 071
  [arXiv:1108.0595 [hep-ph]].

\bibitem{arXiv:0811.0011}
  S.~Kraml, A.~R.~Raklev and M.~J.~White,
  Phys.\ Lett.\ B\ {\bf 672} (2009) 361
  [arXiv:0811.0011 [hep-ph]].

\bibitem{Barger:2010aq}
  V.~Barger, G.~Shaughnessy and B.~Yencho,
  Phys.\ Rev.\ D {\bf 83} (2011) 055006
  [arXiv:1011.3526 [hep-ph]].
  
\end{thebibliography}
\end{document}